# Leakage-Aware Interconnect for On-Chip Network

Yuh-Fang Tsai, Vijaykrishnan Narayaynan, Yuan Xie, and Mary Jane Irwin

Computer Science and Engineering, Penn State University

## ABSTRACT

*On-chip networks have been proposed as the interconnect fabric for future systems-on-chip and multi-processors on chip. Power is one of the main constraints of these systems and interconnect consumes a significant portion of the power budget. In this paper, we propose four leakage-aware interconnect schemes .Our schemes achieve 10.13%~63.57% active leakage savings and 12.35%~95.96% standby leakage savings across schemes while the delay penalty ranges from 0% to 4.69%.*

## 1. INTRODUCTION

While Network-on-Chip (NoC) is becoming an attractive alternative to the traditional global interconnect structure to address the design challenges of future high-performance nanoscale architectures, power consumption remains a significant constraint. In the deep sub-micron era, the interconnect wires and associated driver circuits consume an increasing fraction of the energy budget of the system. Given that there are plenty of techniques proposed to reduce the leakage power in memory structures such as buffers and the effectiveness of applying those techniques in router buffers have been proven [1], we focus on low leakage crossbar designs in this work. In [2], bus encoding technique is developed so that leakage reduction is achieved through selective use of high threshold voltage transistors in the buffers (staggered threshold buffers). In this paper, we apply similar staggered threshold voltage buffers in the crossbar designs.

The rest of the paper is organized as follows. In the next section, the proposed leakage-aware interconnect designs and microarchitectural enhancements on the schemes are presented. The evaluations and experimental results are in Section 3 while the conclusions in Section 4.

## 2. LEAKAGE-AWARE INTERCONNECT

### 2.1 Daul-$V_t$ Feedback Crossbar (*DFC*)

Fig. 1 shows the dual-Vt feedback crossbar (*DFC*) switch with output driver of one output port. To manage the leakage power in standby mode, a sleep transistor $N_5$ which is controlled by sleep signal is added. This sleep transistor is shared by all the bits in a flit and it incurs negligible area overhead since wires dominate the area. While a router will be idle for a given amount of idle time, the sleep signal is set to HIGH and node *A* is pulled to GND level reducing the gate leakage of the pass transistors ($N_1$-$N_4$).

### 2.2 Dual-$V_t$ Pre-Charged Crossbar (*DPC*)

The main idea of *DPC* is to have interconnects that have smaller delay time for one polarity of data than the other. By doing so, the output driver can be designed with asymmetric $V_t$ transistors to favor the speed of the other polarity of data. To achieve this, we pre-charge the output wire to a predefined state, $V_{dd}$ in our example, so that it has virtually zero delay time for data in logic 1 state. To balance the delay time, instead of sizing the inverters in drivers asymmetrically, we use asymmetric-$V_t$ leakage–aware inverters. Fig. 2 shows an example of the output to PE port of one case where a node is pre-charged to HIGH in *DPC*. A simple implementation is to pre-charge the output wires by transistor $P_1$ in the negative phase of the clock signal eliminating the delay penalty for low to high transition. When there is no request sent to the arbiter from all the input buffers or when in sleep mode, the pre-charge signal (*pre*) is deactivated to prevent switching power penalties.

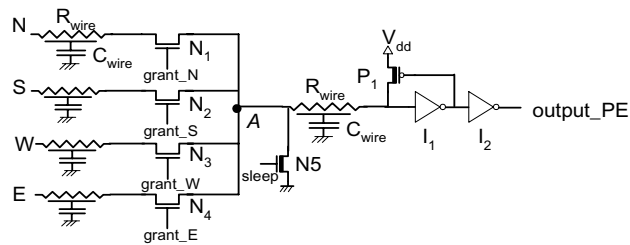

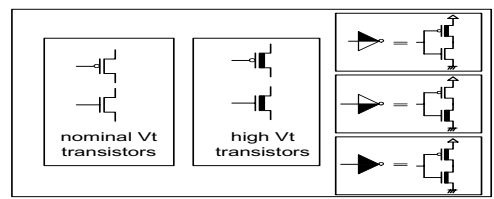

**Figure 1. Schematic of output to PE direction path of DFC.**

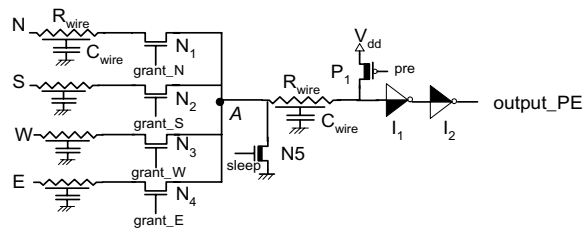

**Figure 2. Output to PE path of pre-charged to HIGH *DPC* design.**

### 2.3 Segmented Dual-$V_t$ Feedback Crossbar (*SDFC*)

Fig. 3(a) shows the segmented dual-Vt feedback crossbar (*SDFC*). As can be seen, the path 1 (bold solid line) has smaller capacitive and resistive loads as compared to path 2 (dashed line). This provides longer slack for transitions through path 1 than that through path 2. The longer slack removes more transistors from the critical path, allowing designers to use high $V_t$ transistors. This scheme not only adds more high Vt transistors but also results in higher probability that some segments of the wires can be put in standby mode.

### 2.4 Segmented Dual-$V_t$ Pre-Charged Crossbar (*SDPC*)

An example of the pre-charged to high *SDPC* is shown in Fig. 3(b). The longer slack in the paths in the shaded area allows all transistors in their output drivers to be of high $V_t$. Moreover, the use of pre-charge transistors eliminates the threshold voltage drop limitation of using pass transistors and thus no level restoration requirement.

This work was supported in part by MARCO/DARPA GSRC Grant, NSF 0093085, and NSF 0202007.

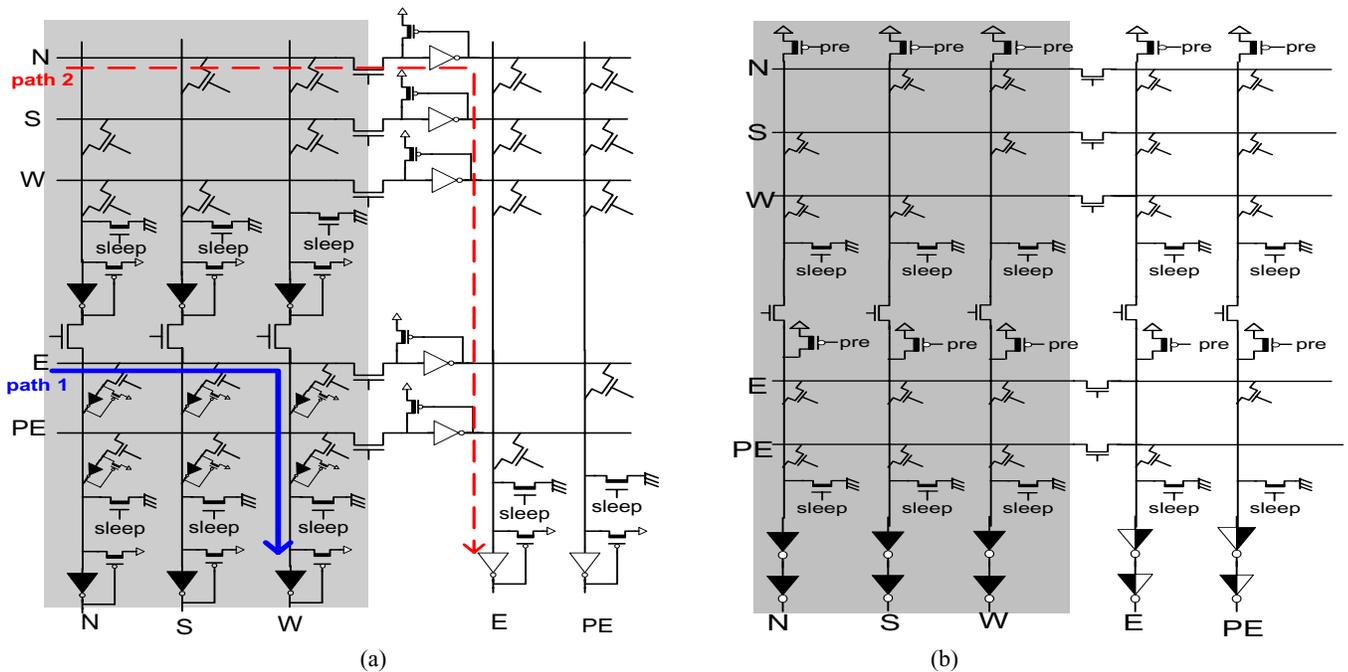

**Figure 3.** Segmented dual-$V_t$ (a) feedback crossbar (b) pre-charged crossbar design.

**Table 1.** Summary of simulation results for evaluated crossbar schemes. Note: * The power consumptions are obtained by assuming 50% static probability which is the worst case for power. **The savings and penalty are the results compared against results of *SC*.

|  | SC | DFC | DPC | SDFC | SDPC |
|---|---|---|---|---|---|
| High to low delay time (ps) | 61.40 | 51.87 | 53.08 | 62.81 | 54.90 |
| Low to High / Precharge delay time (ps) | 54.87 | 58.17 | 61.25 | 64.28 | 62.80 |
| Active Leakage Savings ** | - | 10.13% | 43.7% | 42.09% | 63.57% |
| Standby Leakage Savings ** | - | 12.36% | 93.68% | 43.91% | 95.96% |
| Minimum Idle Time – 3GHz (cycles) | 3 | 2 | 1 | 3 | 1 |
| Total Power – 3GHz (mW) ** | 182.81 | 154.07 | 180.45* | 122.18 | 168.55* |
| Delay Penalty ** | - | No | No | 4.69% | 2.28% |

## 3. EXPERIMENTS AND RESULTS

We implemented the proposed schemes for a 5-by-5 matrix crossbar design with 128 bits per flit in 45nm technology. The interconnect properties, such as wire pitch, space, aspect ratio, and dielectric material parameters, are based on the ITRS roadmap [3]. We predict the interconnect resistance and capacitance by the interconnect model of Berkeley Predictive Technology Model (BPTM) [4]. Except the proposed schemes, the scheme *SC*, whose circuit is the same as the *DFC* except for using a single nominal $V_t$, is also implemented as the base case.

Simulation results are summarized in Table 1. It is intuitive that the inclusion of high $V_t$ device saves both active and standby leakage power. As compared to *DFC* and *SDFC*, *DPC* and *SPDC* offer significantly higher standby leakage savings which are 93.68% and 95.96%, respectively. This is because their output drivers are in the resulting minimum leakage state in standby mode in *DPC* and *SDPC*. Switching to standby mode incurs a switching power penalty, however. We define the *Minimum Idle Time* shown in Table 1 as the minimum amount of time that a circuit stays in idle so that the leakage saved in standby mode is more than the switching power penalty. We would like to note that by segmenting the crossbar, not only is dynamic power mitigated but the leakage power is further reduced by 20% and 30% in *SDFC* and *SDPC*, respectively. This is achieved by our microarchitectural improvement in the output driver designs. Besides, segmentation also increases the chance for a segment to be put in the standby mode for maximum leakage savings.

## 4. CONCLUSIONS

As existing interconnect designs in on-chip network draws significant leakage power, we proposed several dual-$V_t$ designs to reduce both active and standby leakage. The *DFC* saves 10.13% of active leakage and 12.36% of standby leakage while *DPC* saves 43.7% of active leakage and 93.68% of standby leakage at no delay penalty. Optimizing the interconnect structure by segmenting the interconnect and properly assigning the high $V_t$ transistors reduce the leakage further by 30% more in *SDFC*. *DPC* and *SDPC* target to systems which have major data transfers within the same polarity.

## REFERENCES

[1] Chen, X. and Peh, L.-S., "Leakage Power Modeling and Optimization in Interconnection Networks", ISLPED 2003, p. 90-95

[2] H. Deogun, R. Rao, D. Sylvester, & D. Blaauw. "Leakage- and Crosstalk-Aware Bus Encoding for Total Power Reduction". IEEE/ACM DAC 2004, p. 779-782

[3] International Technology Roadmap for Semiconductors. http://public.itrs.net

[4] Berkeley Predictive Technology Model and BSIM4. http://www-device.eecs.berkeley.edu/research.html